\documentstyle[epsf]{mn}
\newif\ifAMStwofonts

\newcommand{\bea}{\begin{eqnarray}}
\newcommand{\eea}{\end{eqnarray}}
\def\lt{\left}
\def\rt{\right}

\def\gte{\,\lower.6ex\hbox{$\buildrel >\over \sim$} \, }
\def\lte{\,\lower.6ex\hbox{$\buildrel <\over \sim$} \, }

\def\be{\begin{equation}}
\def\ee{\end{equation}}
\def\ben{\begin{eqnarray}}
\def\een{\end{eqnarray}}

\def\ba{\begin{eqnarray}}
\def\ea{\end{eqnarray}}

\def\negenspace{\kern-1.1em}

\def \quabla{\sqcup \kern-9pt \sqcap}

\ifoldfss
  \ifCUPmtlplainloaded \else
    \NewTextAlphabet{textbfit} {cmbxti10} {}
    \NewTextAlphabet{textbfss} {cmssbx10} {}
    \NewMathAlphabet{mathbfit} {cmbxti10} {} 
    \NewMathAlphabet{mathbfss} {cmssbx10} {} 
  \fi
  \ifAMStwofonts
    \ifCUPmtlplainloaded \else
      \NewSymbolFont{upmath} {eurm10}
      \NewSymbolFont{AMSa} {msam10}
      \NewMathSymbol{\upi}     {0}{upmath}{19}
      \NewMathSymbol{\umu}     {0}{upmath}{16}
      \NewMathSymbol{\upartial}{0}{upmath}{40}
      \NewMathSymbol{\leqslant}{3}{AMSa}{36}
      \NewMathSymbol{\geqslant}{3}{AMSa}{3E}

      \let\leq=\leqslant 
      \let\geq=\geqslant 
    \fi
  \fi
\fi 

\ifnfssone
  \newmathalphabet{\mathit}
  \addtoversion{normal}{\mathit}{cmr}{m}{it}
  \addtoversion{bold}{\mathit}{cmr}{bx}{it}
  \newmathalphabet{\mathbfit} 
  \addtoversion{normal}{\mathbfit}{cmr}{bx}{it}
  \addtoversion{bold}{\mathbfit}{cmr}{bx}{it}
  \newmathalphabet{\mathbfss} 
  \addtoversion{normal}{\mathbfss}{cmss}{bx}{n}
  \addtoversion{bold}{\mathbfss}{cmss}{bx}{n}
  \ifAMStwofonts
    \ifCUPmtlplainloaded \else
      \UseAMStwoboldmath
      \makeatletter
      \new@mathgroup\upmath@group
      \define@mathgroup\mv@normal\upmath@group{eur}{m}{n}
      \define@mathgroup\mv@bold\upmath@group{eur}{b}{n}
      \edef\UPM{\hexnumber\upmath@group}
      \new@mathgroup\amsa@group
      \define@mathgroup\mv@normal\amsa@group{msa}{m}{n}
      \define@mathgroup\mv@bold\amsa@group{msa}{m}{n}
      \edef\AMSa{\hexnumber\amsa@group}
      \makeatother
      \mathchardef\upi="0\UPM19
      \mathchardef\umu="0\UPM16
      \mathchardef\upartial="0\UPM40
      \mathchardef\leqslant="3\AMSa36
      \mathchardef\geqslant="3\AMSa3E

      \let\leq=\leqslant 
      \let\geq=\geqslant 
    \fi
  \fi
\fi 

\ifnfsstwo
  \DeclareMathAlphabet{\mathbfit}{OT1}{cmr}{bx}{it}
  \SetMathAlphabet\mathbfit{bold}{OT1}{cmr}{bx}{it}
  \DeclareMathAlphabet{\mathbfss}{OT1}{cmss}{bx}{n}
  \SetMathAlphabet\mathbfss{bold}{OT1}{cmss}{bx}{n}
  \ifAMStwofonts
    \ifCUPmtlplainloaded \else
      \DeclareSymbolFont{UPM}{U}{eur}{m}{n}
      \SetSymbolFont{UPM}{bold}{U}{eur}{b}{n}
      \DeclareSymbolFont{AMSa}{U}{msa}{m}{n}
      \DeclareMathSymbol{\upi}{0}{UPM}{"19}
      \DeclareMathSymbol{\umu}{0}{UPM}{"16}
      \DeclareMathSymbol{\upartial}{0}{UPM}{"40}
      \DeclareMathSymbol{\leqslant}{3}{AMSa}{"36}
      \DeclareMathSymbol{\geqslant}{3}{AMSa}{"3E}

      \let\leq=\leqslant 
      \let\geq=\geqslant 
    \fi
  \fi
\fi 

\ifCUPmtlplainloaded \else
  \ifAMStwofonts \else 
    \def\upi{\pi}
    \def\umu{\mu}
    \def\upartial{\partial}
  \fi
\fi

\title{Scalar field haloes as gravitational lenses}
\author[F.E. Schunck, B. Fuchs \& E.W. Mielke]
       {Franz E. Schunck$^1$, Burkhard Fuchs$^2$ \&
       Eckehard W. Mielke$^3$\\
$^1$Institut f\"ur Theoretische Physik, Universit\"at zu K\"oln,
50923 K\"oln, Germany\\
        $^2$Astronomisches Rechen--Institut, M\"onchhofstr.~12--14, 69120
        Heidelberg, Germany,\\ $^3$Departamento de F\'{\i}sica,
         Universidad Aut\'onoma Metropolitana--Iztapalapa,
       Apartado Postal 55-534, C.P. 09340, M\'exico, D.F., M\'exico}
\date{Accepted
      Received ;
      in original form 2005}

\pagerange{\pageref{firstpage}--\pageref{lastpage}}
\pubyear{2006}

\begin{document}

\maketitle

\label{firstpage}

\begin{abstract}
A non-topological soliton  model with a repulsive scalar self-interaction of 
the Emden type provides a constant density core,
similarly as the empirical Burkert profile of 
dark matter haloes. As a further test, we derive the gravitational lens 
properties of our model, in particular, the demarcation curves 
between `weak' and `strong' lensing. Accordingly,  strong 
lensing with typically three images is almost three times more probable 
for our solitonic model than for the Burkert fit. Moreover, some prospective 
consequences of a possible flattening of dark matter haloes are indicated. 
\end{abstract}

\section{Introduction}
Real scalar fields could provide
mutual connections between cosmology and particle physics.
The first attempt to admit a gravity scalar was by Nordstr\o m (1913), 
whereas Jordan (1959)
as well as Brans \& Dicke (1961) introduced it as a means for  varying the 
gravitational `constant' $G$ with time.
The most intensive application of  scalar fields occurs in  the early universe
scenarios. A prime example is inflation  driven
by the potential energy of the real scalar field, the inflaton, and in
phase transitions  together with the production of topological defects
(Vilenkin \& Shellard 1994). It  also arises in the theories
of superstrings where the
dilaton couples universality to the trace of the energy-momentum tensor
 and leads to the so-called
superinflation driven by the kinetic energy of the scalar field
(see ~Gasperini \& Veneziano 1993a, 1993b).

One of the main predictions of general relativity (GR) is the
{\em deflection of
light} in gravitational fields of compact objects known as {\em gravitational
lensing}. Although the deflection of light by the Sun, up to a  factor 
{\em of two}, was  suggested    by
Soldner already in 1801  in the context of Newton's theory,
the correct value of
$\delta =4 M_\odot/R_\odot \simeq 1,75^{\prime\prime}$  was derived within the
framework of general
relativity by Chwolson (1924) and Einstein (1936),  see
Schneider et al. (1992) for more details. Intensive studies of gravitationally 
lensed objects began  with the observation (Young et
al. 1980) of the double quasar 
 Q0957+561 with redshift  $z=0.39$ and were mainly concentrated on the
weak gravitational lensing effect for which the  deflection is
very small, up to a few arcseconds (see
 Bartelmann \& Schneider 2001 for a recent review). In general,
one can distinguish between lensing on non-cosmological and cosmological
distances. As for the former one can have lensing by a star of solar mass
$M_\odot$ in a galaxy with the deflection angles of the order
of milliarcseconds and by a galaxy of $10^{11} M_\odot$ with the
deflection angles of the order of arcseconds.
Although model-dependent luminosity distances have to be assumed,  this 
provides us with the opportunity to
determine the mass distribution of dark matter haloes, in particular the inner 
density slopes. Moreover, information on the cosmological parameters such as 
the Hubble constant $H_0$,
the deceleration parameter $q_0$ and the cosmological constant $\Lambda$
(Fukugita et al.~1990), can be obtained.

In the case of weak lensing, a comparison with Newton's theory
is instructive, but all the Newtonian-based intuition fails
for the cases of the
large deflection angles. If one has a black hole, or a strong compact object
such as a neutron star, then the gravitational lensing effects have to
be treated in fully relativistic terms (Petters 2003). The deflection angles 
can be of the order of  degrees allowing the light to orbit the lens several
times such that  multiple images can be formed (Cramer 1997, Dabrowski \& 
Schunck 2000). These effects are
especially interesting for accretion disks (Peitz \& Appl 1997).
It is worth mentioning the  possibility of
gravitational lensing by exotic matter such as cosmic strings
(Vilenkin \& Shellard 1994), boson stars (Schunck \& Mielke, 1999, 2003)
and by gravitational waves (Faraoni 1993).

Recently, Virbhadra et al.~(1998, 2002) coupled a real
massless scalar field to the Einstein equations and got a general static
spherically symmetric asymptotically flat solution with a naked
singularity which, apart from mass, has an extra parameter 
called ``scalar charge''. They found a possibility of having four
images of a source at the observer position together with
a double {\em Einstein ring}.

In this paper, we investigate the gravitational lensing of a 
spherically symmetric  halo  constructed from an exactly solvable scalar field 
model with a
{\em self-interaction} of the Emden type. To some extent this adopts  the
proposal of Spergel  \& Steinhardt (2000) that dark matter (DM) could be  
weakly self-interacting, or even a Bose-Einstein condensate
(Goodman, 2000) with a repulsive potential. Then, DM would
be more `collisional' and may avoid the central cusps of the conventional
cold dark matter (CDM) haloes.

The paper is organized as follows. In Section 2 we study the photon 
trajectories  arising from null geodesics in the gravitational field of a 
spherically symmetric halo. In Section 3 we derive 
the density profile of our non-topological soliton (NTS) model and compare it 
with other empirical profiles, such as the quasi-isothermal sphere or the 
Burkert fit. 
The rotation curves derived for the NTS halo in Section 4 are contrasted 
to those emerging from the isothermal sphere and 
the Burkert fit. In Section 5, we  
discuss the effects of gravitational lensing by such bosonic configurations,  
calculate exactly the normalized projected mass, corrections due to pressure,
and draw  the dimensionless 
weak-field lens equation. Demarcation curves between `weak' and `strong' 
lensing  for the NTS halo are compared with those of other prominent models in 
Section 6.  The location of Einstein rings, the image separation and 
magnification
are provided in Section 7.  Section 8 summarizes
the results and gives an outlook on the consequences of 
a possible flattening of NTS haloes.

\section{Photon trajectories in a gravitational field}

For reasons mentioned in the introduction gravitational lensing requires 
a full {\em general-relativistic framework}:

For spherically symmetric halo, the line element reads
\be
ds^2 = e^{\nu (r)} dt^2 - e^{\tilde\lambda (r)} dr^2 - r^2 \left(
 d\vartheta^2 + \sin^2\vartheta d\varphi^2 \right)  \; ,  \label{met}
\ee
in which the functions $\nu =\nu (r)$ and
$\tilde\lambda =\tilde\lambda (r)$ depend on the Schwarzschild type radial
coordinate $r$.

The main point of our analysis is the gravitational lensing properties
by a bosonic halo and this requires the discussion of trajectories of photons
along a null geodesics  in the plain $\vartheta =\pi/2$. It is equivalent to 
the energy conservation
\be
\frac{1}{2} \left(\frac{dr}{d\tau}\right)^2 + V_{\rm eff}  =
  \frac{1}{2} F^2 e^{-\tilde\lambda -\nu}
\label{null1}
\ee
of a particle with kinetic energy moving in the effective gravitational 
potential
\be
V_{\rm eff} = e^{-\tilde\lambda } \frac{L^2}{2 r^2} \; . \label{veff}
\ee
Here $F:=e^\nu dt/d\tau$ is the total energy, $L:=r^2 d\varphi/d\tau$
the angular momentum of the photon, and
$\tau$  an affine parameter along null
geodesics, such that 
$L/F= r^2e^{-\nu} d\varphi/dt= re^{-\nu/2}$. 

When the light travelling from a source is
deflected by a halo,  the deflection angle is given by
(Sexl \& Urbantke 1983, p. 110)
\be
\hat{\alpha}(r_0) =2 \int_{r_0}^\infty
\frac{be^{\tilde\lambda /2}}{\sqrt{r^2 e^{-\nu} - b^2}} \frac{dr}{r} - \pi \; ,
\label{exactbend}
\ee
where
\be
b=L/F(r_0)=r_0\exp{[-\nu(r_0)/2]} \; ,
\ee
is the impact parameter and and $r_0$ the closest distance between a light ray
and the center
of the halo for which
$V_{\rm eff}(r_0) = (F^2/2) \exp[-\nu(r_0) - \tilde\lambda(r_0)]$ holds.

The diagonal components of the energy-momentum tensor
$T_\mu{} ^\nu(\Phi) = {\rm diag} \; (\rho , -p_r,$
$-p_\bot, -p_\bot )$ of a  scalar field are
\ben
\rho &=& \frac{1}{2} \left ( \omega^2 P^2 e^{-\nu} +P'^2 e^{-\tilde\lambda} + U
 \right ) \; , \nonumber  \\
p_{\rm r} &=&  \rho -  U \; , \;  \nonumber  \\
p_\bot &=&  p_r -  P'^2 e^{-\tilde\lambda } \; . \label{emt}
\een
which, due to  $\nabla^{\mu} T_{\mu}{}^{\nu}= 0$, satisfy the {\em generalisation} 
\be
 \frac{d}{dr} p_{\rm r}= -\nu^\prime\left(\rho + p_{\rm r} -\frac{2}{r}(
p_{\rm
 r}-p_\bot)\right) \label{TOV}
\ee
 of the
Tolman-Oppenheimer-Volkoff equation for 
an anisotropic `fluid',  cf. Schunck \& Mielke (2003).

{}For  stationary configurations,
 the equivalent radial Klein-Gordon equation reads
\be
P'' + \left( \frac {\nu' - \tilde\lambda'}{2} + \frac {2}{r} \right)
 P' =e^{\tilde\lambda}  \left[ \frac{dU(P)}{dP^2} -e^{- \nu }
\omega^2\right] P\,, \label{2ska}
\ee
where $P:= P(r)=\Phi (r,t) e^{i\omega t}$ is the radial function and $U$ a 
self-interaction.
 
The decisive non-vanishing components of the Einstein equation are
the `radial' equations
\ben
\nu' + \tilde\lambda' & = & \kappa (\rho + p_{\rm r}) r e^{\tilde\lambda}
\; ,  \label{nula}\\
\tilde\lambda' & = & \kappa \rho r e^{\tilde\lambda} - \frac {1}{r} 
(e^{\tilde\lambda} - 1)
\; , \label{la}
\een
where $\kappa= 8\pi G$ is the gravitational coupling constant in natural units.
Two further  components are identically fulfilled because of
the contracted Bianchi identity $\nabla^{\mu} G_{\mu}{}^{\nu}\equiv 0$ and 
the energy-momentum conservation(\ref{TOV}).

One of the metric functions has the well-known exact parametric solution
\be
e^{-\tilde\lambda(r)} = 1 - \frac{2GM(r)}{r}= 1 - \frac{\kappa M(r)}{4\pi r} \; ,
\label{paralam}
\ee
where
\be
M(r) := 4\pi\int_0^r \rho {\tilde r}^2 d{\tilde r}  \label{Newmass}
\ee
is the Newtonian mass function at the distance $r$ from the center.

\section{Density profile of non-topological solitons
  \label{ntssol}}

As a solvable toy model (Mielke, 1978, 1979) with {\em self-interaction}, let 
us consider the Klein-Gordon equation (\ref{2ska}) with
a $\Phi^6$ type potential
\be
U(\vert  \Phi \vert ) = m^2 \vert  \Phi \vert ^2 \left(1 -
\chi \vert  \Phi \vert^4 \right), \qquad \chi \vert  \Phi \vert^4\leq 1\,,
\label{pot6}
\ee
where $m$ is the `bare' mass of the boson and  $\chi$
a coupling constant, which are thought of as constants of nature, cf.
Mielke \& Schunck (2002), Mielke, et al. (2002).

For a spherically symmetric configuration and
the choice $\omega =m$, the corresponding
nonlinear Klein-Gordon equation (\ref{2ska}) simplifies in the limit 
$\nu=\tilde\lambda\rightarrow 0$  of 
flat spacetime to an {\em Emden type equation}
\be
P^{\prime\prime} + \frac{2}{x} P^{\prime}
  + 3\chi P^5 = 0\, ,
\label{Emden}
\ee
which has the completely {\em regular} exact solution
\be
P(r) = \pm \chi^{-1/4} \sqrt{\frac{A }{1+A^2 x^2}} \, , \label{NTS}
\ee
where  we introduced
the {\em dimensionless} radial coordinate $x:=m r$,
and $A = \sqrt{\chi} P^2(0)$ is a free intergration constant depending 
on the central value $P(0)$.
As is rather
characteristic  for {\em non-topological soliton} (NTS)
solutions, its dependence on
the nonlinear coupling parameter $\chi$ is singular in the limit
$\chi\rightarrow 0$.

{}From (\ref{emt}) we find in flat spacetime the energy density
\ben
\rho 
 &=& \frac{2\rho_0}{2-A^2}
 \left[\frac{ r_{\rm c}^2}{r_{\rm c}^2+r^2} +
 \frac{A^2 r_{\rm c}^4( r^2- r_{\rm c}^2)}{2(r_{\rm c}^2+r^2)^3}\right]
   \; ,
\label{rhonts}
\een
where the central density of the NTS model
$\rho_0 =A(2-A^2)m^2/2\sqrt{\chi}$ is positive for $A < \sqrt{2}$
and the core or {\em scale} radius is $r_{\rm c}=1/m A$.
{}For small $A$, the leading term of the Newtonian type mass
concentration (\ref{rhonts})
is exactly the density law of the {\em quasi--isothermal sphere}  (IS)
\begin{equation}
\rho_{\rm IS} (r) = \rho_0
\frac{ r_{\rm c}^2}{r_{\rm c}^2+r^2} \; .
\label{quasirho}
\end{equation}
At large radii the density falls of like $\rho \propto r^{-2}$ which
corresponds to an asymptotically flat rotation curve.
The NTS density (\ref{rhonts}) implies a {\em scaling law} for the dark haloes
of the form
\begin{equation}
\rho_0 = \frac{m}{\sqrt{\chi}} \frac{1}{r_{\rm c}}
\left(1 - \frac{1}{2m^2r_{\rm c}^2}\right)
\simeq \frac{m}{\sqrt{\chi}}\frac{1}{r_{\rm c}}\propto \frac{1}
{r_{\rm c}} \; ,  \label{scallaw}
\end{equation}
where $A$, which may vary from halo to halo,
{\em cancels out}. The remaining quotient $m/\sqrt{\chi}$ may be tentatively 
regarded as universal constant of scalar dark matter (DM).

For $r_{\rm c}>>1/m$, which is equivalently to $\chi   P^4(0)<< 1$
and therefore within the positive range of
the NTS self-interaction (\ref{pot6}), the approximate scaling relation
$\rho_0 \propto 1/
{r_{\rm c}}$ has been successfully tested by
Fuchs \& Mielke (2004) against the observed properties of  various types of 
galaxies.

Let us contrast these results with the density law of 
Navarro, Frenk \& White (1996) based on  simulations of conventional, 
i.e., non-interacting CDM:
The corresponding generalized NFW profile
 \begin{equation}
\rho_{\rm NFW} (r) = \rho_0
\frac{ r_{\rm c}^3}{r^\epsilon(r_{\rm c}+r)^{3-\epsilon}}\; ,
\quad 0<\epsilon <3 .
\label{NFWrho}
\end{equation}
is, however,  singular with a cusp at the origin. The original NFW profile 
corresponds to $\epsilon=1$. More realistic appears to be the empirical density
profile
\begin{equation}
\rho_{\rm B} (r) = \rho_0
\frac{ r_{\rm c}^3}{(r_{\rm c}+r)(r_{\rm c}^2+r^2)}  \; .
\label{Brho}
\end{equation}
of Burkert (1995).
It has a constant density core, similarly as the NTS halo, but
a $\rho \propto r^{-3}$ fall-off at infinity, similarly as the NFW
profile.

\section{Rotation curves}

In general, for the static spherically symmetric metric (\ref{met}),
an observer at rest at the equator of the  Schwarzschild
type
coordinate system measures the following
tangential velocity squared as a `point particle'  (a star or interstellar HI 
gas) flies past him in its circular orbit,
{\em cf.} Misner {\em et al.} (1973), p.~657, Eq.~(25.20):
\ben
v^2_\varphi &:=& e^{-\nu} r^2 \left ( \frac{d\varphi}{dt} \right )^2
= \frac{1}{2} r\nu^\prime
= \frac{1}{2}\left[1 -
 e^{-\tilde\lambda}  +\kappa p_{\rm r}r^2 \right]e^{\tilde\lambda }\nonumber\\
&\simeq& \frac{\kappa}{2}\left[\frac{M(r)}{4\pi r} + p_{\rm r}r^2\right]
 \, . \label{tanvel}
\een

As is well-known (Wald 1984), a naive application of the Newtonian limit
would have led us to geodesics in flat spacetime, i.e.~as if
gravity would not affect the motion of test bodies like our stars moving
in the DM halo. Thus it is mandatory to go beyond, as is
indicated by our approximation of the generally
relativistic formula (\ref{tanvel}).
Then,  also the  pressure component $p_r \neq 0$ of an anisotropic `fluid'
contributes, as is the case of scalar
fields.

As our simplest fiducial example, we calculate the total mass 
\be
M(r)= 4 \pi \rho_0 r_{\rm c}^3 [ r/r_{\rm c} - \arctan(r/r_{\rm c}) ] \, ,
\ee
of the IS halo (\ref{quasirho}) which implies  the corresponding  circular 
velocity
\be
v_\varphi^2= \frac{\kappa}{2}  \rho_0 r_{\rm c}^2 \left[1 -
\frac{r_{\rm c}}{r}\arctan\left(\frac{r}{r_{\rm c}}\right)\right ] \, .
\ee

On the other hand, from the Newtonian NTS solution (\ref{NTS}) and its radial
pressure (\ref{emt}), we find  the related rotation velocity
\be
v^2_\varphi/ v^2_\infty =
 1 + \left(\frac{A ^2}{8}-1\right)\frac{r_{\rm c}}{r}
     \arctan\left(\frac{r}{r_{\rm c}}\right)
 + \frac{A ^2 r_{\rm c}^2(r^2-r_{\rm c}^2)}{8(r_{\rm c}^2+r^2 )^2},
   \label{NTSrot}
\ee
where
\be
 v^2_\infty =
 \frac{\kappa\rho_0 r_{\rm c}^2}{2-A^2} \leq 10^{-6} \label{vinfty}
\ee
is restricted by observations.

According to Eq. (\ref{NTSrot}) of the NTS model,
the radial component $e^{- \tilde\lambda }$ of the  metric (\ref{met}) asymptotically
approaches the value
$1-2v^2_\infty<1$. After   
 a redefinition of the
radial coordinate $r \rightarrow \tilde r:= r/\sqrt{1-2v^2_\infty}$,  
the asymptotic space  has a {\em  solid deficit angle}: The area of a sphere
of radius $r$ is not $4 \pi r^2$, but $4 \pi (1-2v^2_\infty) r^2$. However, 
there is a remedy by identifying the boundaries of the deficit angle and 
thereby retaining an asymptotically flat metric, at the price of a conical 
singularity (Tod 1994) at 
the origin, similarly as in the case of cosmic strings, global monopoles
and global textures, cf.  Vilenkin \& Shellard (1994).

The more {\em realistic} phenomenological
Burkert fit (\ref{Brho})   amounts
asymptotically  to a {\em  logarithmic modification} of the
Kepler law such that the velocity tends to zero at
spatial infinity, with the consequence that
{\em no} such deficit angle is to be expected. 

\section{Gravitational lensing by dark matter haloes}
The lens equation {\em for large} deflection angles may be expressed as
(Virbhadra \& Ellis  2000)
\be
\tan\beta = \tan\theta -
\frac{D_{\rm ls }}{D_{\rm  s }}
\left[\tan{\theta} + \tan(\hat{\alpha} - \theta) \right]
\ee
where $D_{\rm ls }$ and $D_{\rm  s }$ are distances from the
lens (deflector) to the
source and from the observer to the source, respectively,  $\beta$ denotes
the true angular position
of the source, whereas $\theta$ stands for the image positions.
Observe that, for large angles, the distance $D_{\rm ls }$
cannot be considered a constant but it is a function of the
deflection angle (Dabrowski \& Schunck 2000).

One usually defines the reduced deflection angle to be
\be
\alpha := \theta - \beta = \sin^{-1}\lt(\frac{D_{\rm ls }}{D_{\rm  s }}
\sin\hat{\alpha}\rt) \, .  \label{redang}
\ee

Since the gravitational potential $\vert\phi\vert< 10^{-4}$c$^2$ of a galaxy 
is rather weak and neglecting pressure, one can use the linear approximation 
to Einstein's equations, for which 
\be
e^\nu \simeq 1 +2 \phi= 2-n\,, \quad   e^{\tilde\lambda} \simeq 1 +2r \phi^\prime
\label{weakrefrac}
\ee 
holds in radial coordinates. Here $n:= 1-2\phi/c^2$ can be regarded
as an effective refraction index in flat spacetime.
 
Then the Newtonian potential 
\be
\phi= -\frac{\kappa}{8\pi}  \int
   \frac{\rho ({\vec r}\;^\prime)}{|{\vec r}-
   {\vec r}\;^\prime|}
  d^3 {r}^\prime  \; ,
\ee
as a solution of  the Poisson equation via the method of Green's functions, 
is only determined by the energy density $\rho$.
Accordingly, in this approximation the 
 deflection angle is given by 
\be
{\vec {\hat{\bf \alpha}}} ({\vec r})=-\int_0^s \vec\nabla_\bot n dl= 
2 \int_0^s \vec\nabla_\bot \phi dl \,, \label{bendappr}
\ee
where $ dl \simeq dz$ is along the light path in an appropriately
choosen cordinate system. {}For a geometrically-thin 
lens  (Narayan \& Bartelmann 1996), one usually considers the mass
sheet orthogonal to the line-of-sight,
commonly called the lens plane. It is characterized by
its {\em surface mass} density
\be
\Sigma ({\vec r}_\bot) = \int \rho({\vec r}_\bot, z) dz \; , \label{surf}
\ee
where ${\vec r}_\bot$ is a two-dimensional vector in the lens plane. The 
deflection angle at position
${\vec r}_\bot$ is then the sum of the deflections due to all mass elements
in the plane
\be
{\vec {\hat{\bf \alpha}}} ({\vec r}_\bot) =\frac{\kappa}{2\pi} \int
   \frac{({\vec r}
 - {\vec r}\;^\prime)_\bot \Sigma ({\vec r}_\bot\;^\prime)}{|{\vec r}-
   {\vec r}\;^\prime|_\bot^2}
  d^2 {r}_\bot^\prime  \; .
\ee
In general, this deflection angle is a two-dimensional vector.
In case of circular symmetry, the coordinate origin can be shifted to the
center of symmetry and, so, light deflection is reduced to a
one-dimensional problem.

Then the {\em dimensionless weak-field lens equation} can be written as
\be
\ \beta( \theta) = \theta - \lambda
\frac{g(\theta)}{\theta}\; ,
\quad \lambda:=  \frac{4\rho_0 r_{\rm c}}{\Sigma_{\rm cr}} \, ,
\label{dimlens}
\ee
where $\theta= \vert {\vec r}_\bot\vert/D_{\rm l}$ and
$\Sigma_{\rm cr}=2D_{\rm  s } /(\kappa D_{\rm l}D_{\rm ls })$ is
the {\em critical} surface
density. The function $g$ is the normalized projected mass
\ba
M_\bot(r_\bot)&:=&
 2\pi\int_0^{r_\bot} du \, u \int_{-\infty}^\infty
  \rho \left ( \sqrt{u^2 +w^2} \right ) dw
 \nonumber \\
&=&
4\pi \rho_0 r_{\rm c}^3 g\left(\frac{r_\bot}{r_{\rm c}}\right)
 \label{nopromass}
\ea
enclosed within a projected radius $r_\bot$ and integrated along
$w:=z/r_{\rm c}$.
An {\em Einstein radius} is defined by the condition
$\beta(\theta_{\rm E}) \equiv 0$,
or $\theta_{\rm E}^2 =\lambda g(\theta_{\rm E})$.

{}For the IS halo, the normalized projected mass is given by
\ba
g_{\rm IS}(x) &=& \int_0^x u du \int_0^\infty \frac{ dw}{1+u^2 +w^2}
   \nonumber\\
 &=&
\frac{\pi}{4}\int_0^x \frac{du^2}{\sqrt{1+u^2}}=
   \frac{\pi}{2}\left(\sqrt{1+x^2}-1\right)\,,
\ea
where we applied for the evaluation of the improper integral the
theorem of residues, and the
 corresponding bending angle is given by 
\be
\alpha_{\rm IS}(x) = \frac{\kappa}{2} \rho_0
 r_{\rm c}^2 \frac{\sqrt{1+x^2} - 1 }{x}\,.
\ee

For the NTS profile, we similarly find  
\ba
g(x)&=&\frac{1}{2-A^2}
 \Bigl[\pi\left(\sqrt{1+x^2}-1\right)\\ &+&
\frac{A^2}{2}\int_0^x du^2\int_0^\infty
  \frac{u^2 +w^2 -1}{(1+u^2 +w^2)^3} dw \Bigr]\nonumber\\
&=& \frac{\pi}{2-A^2} \left[\sqrt{1+x^2}-1 
+ \frac{A^2}{16}\int_0^x   \frac{2u^2 -1}{(1+u^2)^{5/2}}du^2 \right]
   \nonumber\\
&=&\frac{\pi}{2-A^2} \left[\sqrt{1+x^2}-1
  - \frac{A^2}{8} \left ( \frac{1+2x^2}{(1+x^2)^{3/2}} - 1 \right )
  \right]\,.
\nonumber \label{projmaNTS}
\ea

In comparison, for the NFW  profile the projected mass $g(x)$ 
has been calculated by Bartelmann (1996),
Zhang (2004) derived those of its generalizations (\ref{NFWrho}),
whereas Park \& Ferguson (2003),
discuss the weak and strong lensing regimes of the
Burkert profile (\ref{Brho}) for which $\lambda_{\rm cr} =8/\pi$.

\subsection{Corrections due to pressure}
A conserved symmetric energy-momentum tensor $T_{\mu\nu}$ implies for 
 the integrated spatial components  the Laue theorem
\be
 \int T^{AB}d^3 {r} =\frac{1}{2}\frac{d^2}{dt^2} \int T_{00}r^A r^B d^3 {r} 
\ee
in the limit of {\em flat} spacetime, cf. Sexl \& Urbantke (1983, p.~78). Since the
anisotropic stresses (\ref{emt}) of our NTS are {\em static},
they do not contribute to the gravitational potential $\phi$ in this limit.

However, in a higher order approximation, the contribution from the radial pressure 
\be
p_{\rm r} =
\frac{A ^3 m^2}{2\sqrt{\chi}(1+A ^2 x^2)^2} = 
\frac{A ^2\rho_0 }{2- A ^2}\frac{r_{\rm c}^4}{(r_{\rm c}^2 + r^2)^2}
\label{radp}
\ee
of our NTS solution will cause a gravitational potential 
$\widetilde\phi := (e^\nu -1)/2  \simeq\phi +\Psi$, which is 
modified in comparison to the linear  
approximation (\ref{weakrefrac}) valid for configurations without pressure. Since the shift  
 $e^{\tilde\lambda}$ is, according to the exact parametrization (\ref{paralam}), 
only determined by the energy density $\rho$, its 
 weak field approximation remains intact. 
 
The modified gravitational potential $\widetilde\phi$, entering now in the 
corresponding deflection angle (\ref{bendappr}), can readily be obtain by 
 caluculating the  derivative
$\vec\nabla_\bot =2\vec r_\bot \partial/\partial r^2$
perpendicular to the light path, but departing  from the 
{\em general relativistic} relation (\ref{tanvel}) 
for the rotation velocity:
\ba 
\vec\nabla_\bot \widetilde\phi &\simeq&\frac{1}{2}\vec\nabla_\bot (e^\nu -1)= \vec r_\bot r^{-2}
v^2_\varphi e^{\nu}\nonumber\\
&=& 
\frac{\kappa}{2}\vec r_\bot\left[\frac{M(r)}{4\pi r^3}
  + p_{\rm r}\right]e^{\nu + \tilde\lambda} \, ,
\ea
in which the extra contribution from the radial pressure 
$p_{\rm r}$ becomes explicit. 

Since $e^{\nu + \tilde\lambda}\simeq 1$ as  in the linear approximation
and $rdr =\sqrt{r^2 -r_\bot^2}dz$ in
the choosen coordinate system,
we find from Eq.~(\ref{bendappr}) for the  deflection angle of a NTS halo
\be
{\vec {\hat{\bf \alpha}}} ({\vec r}_\bot)=2{\vec r}_\bot\int v^2_\varphi
\frac{d \ln r}{\sqrt{r^2 -r_\bot^2}}\, .
\ee
In cases with a prescribed equation of state, Bharadwaj \& Kar (2003)
have discussed a similar relation between the deflection angle
and the rotation velocity squared.

In this next order approximation, not only  the normalized projected  
mass (\ref{nopromass}) but also the radial pressure $p_{\rm r}$ contributes via
\ba
p(x) &:=& \frac{x^2}{\lambda} \int_{-\infty}^{\infty}p_{\rm r}dz =
\frac{A^2}{2-A^2}\int_0^\infty \frac{x^2 dw}{(1 +x^2 +w^2)^2}
   \nonumber\\
 &=& 
 \frac{\pi A^2}{4 (2-A^2)}
 \frac{x^2}{(1+x^2)^{3/2}}  \nonumber\\
 &\rightarrow &
 \frac{\pi A^2}{4 (2-A^2)} \frac{1}{x} \, ,
\ea
to the bending angle, using
the same normalization. In comparison to $g(x)$, the pressure contribution is
negligible for small as well as large angles.
In fact, the resulting $\widetilde g(x)= g(x) + p(x)$
is similar to the expression (\ref{projmaNTS}),
only the term proportional to $2x^2$ will disappear.

\begin{figure}
\begin{center}
\leavevmode\epsfysize=9cm
      \epsffile{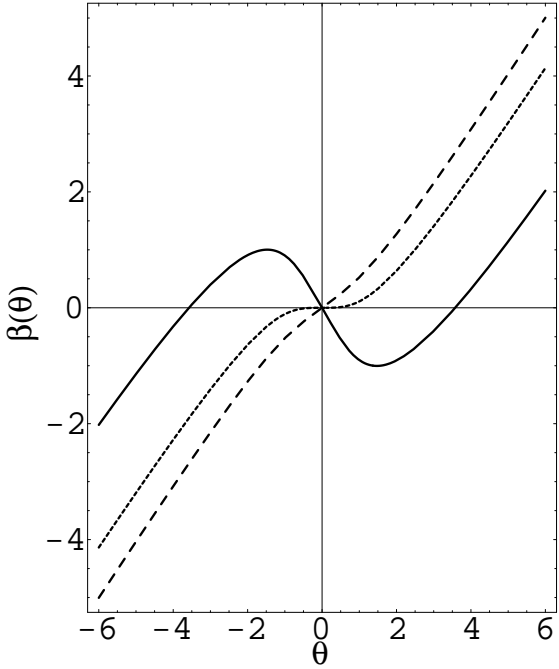}
 \caption{Dimensionless lens equation for the NTS halo 
with presure for $A=0.805$. The solid, dotted, and dashed lines are, respectively,
for $\lambda=2$, $\lambda_{\rm cr}$, $0.3$. Multiple images occur for 
$\lambda>\lambda_{\rm cr}=0.692428$.}
         \label{leseq}
  \end{center}
   \end{figure}

The corresponding lens equation (\ref{dimlens}) is drawn in Fig.~\ref{leseq}
for different $\lambda$'s. Its shape is  similar to that of the Burkert 
profile, cf.  Park \& Ferguson (2003), albeit for a different scale.

Eventually, more precise approximations would lead us to a
general-relativistic treatment 
in which the nonlinear, gravitationally 
coupled KG equation
(\ref{2ska}) together with the full radial
Einstein equations (\ref{nula}) and (\ref{la}) need to be solved
self-consistently, before the deflection
angle (\ref{exactbend}) can be determined.
Since the gravitational potential $\phi$ of a 
galaxy is  weak, we expect rather small corrections to the Emden type 
exact solution (\ref{NTS}) from  
such  numerical simulations, which are beyond the scope 
of the present paper.

Let us also remark  that our model of a {\em self-gravitating} NTS is 
quite different from, e.g.,  
 brane-world scenarios (Harko \& Cheng 2006), in which  corrections 
to the Einstein equations are considered 
due to effective stresses induced by the projected Weyl curvature 
in the hypothetical  bulk. Moreover, contrary to the ad-hoc assumption 
of Pal. et al. (2005), the IS type  density (\ref{rhonts}) for the dark halo 
is an outcome of our NTS model.

\section{Demarcation curves}
Even for our weak field approximation of the gravitational field, one
can distinguish between `weak' and `strong'
lensing, depending whether or not the lense produces one or multiple
images, cf.~Bartelmann \& Schneider (2001). This depends on
the characteristic value of the surface mass density. For the
dimensions lens equation (\ref{dimlens}),
the interval $0< \lambda \leq \lambda_{\rm cr}$ is the range of `weak'
lensing, where the critical value
$\lambda_{\rm cr}$ emerges  from the condition $d \beta/d 
\theta_{\vert_{\theta =0}} =0$ of a saddle point at the origin.
Any lens-source 
combination with $\lambda >\lambda_{\rm cr}$ will produce multiple images.

Let us compare the  critical values  for different profiles:
\be
\lambda_{\rm cr}
=\cases{
0                        &  NFW profile        \cr
16(2-A^2)/[\pi(8+3A^2)]  &  NTS with pressure  \cr
16(2-A^2)/[\pi(8-A^2)]  &  NTS halo           \cr
4/\pi                    &  IS halo            \cr
8/\pi                    &  Burkert profile
}   \label{expect}
\ee

For $0< A < \sqrt{2}$, the critical value of the NTS halo is
larger than that of the NFW halo, but smaller than that of the isothermal
sphere,  assuming for $A\rightarrow \sqrt{2}$ and 
$A\rightarrow 0$  
 the corresponding limits. Consequently, the NTS halo, although
exhibiting no density cusp at the center, can provide stronger lensing than
the Burkert profile, irrerespective of the baryonic component.

Whereas deflection angles of galaxies still seem to be difficult to measure,
{\em demarcation curves} can more easily distinguish between  lensing
scenarios with one or more images.  
Thus these strong demarcation lines could provide  an independent means 
to probe the observational validity
of different density profiles for dark matter haloes.
{}For the NTS halo, 
there is only one image in the range $0<\lambda \leq \lambda_{\rm cr}$, 
beyond $\lambda_{\rm cr}$ gravitational lensing will produce three images.

It is remarkable that {\em constant} $\lambda$ lines, in particular
\be
  \frac{4\rho_0 r_{\rm c}}{\Sigma_{\rm cr}}= \lambda_{\rm cr}\,,
\ee
corresponds to the scaling relation
\be
\rho_0 \propto r_{\rm c}^{-1} \label{scalrel}
\ee
of DM haloes,
which is  parallel to the approximate
scaling relation (\ref{scallaw})  predicted by NTS model. For `maximum disk' 
models, it fits almost ideally 
astronomical observations (Fuchs \& Mielke 2004).
{}For Burkert haloes,  Salucci \& Burkert (2000) proposed
the related  {\em scaling relation}
$\rho_{\rm B}(0) \propto r_{\rm c}^{-2/3}$ on an empirical basis.

\begin{figure}
\begin{center}
\leavevmode\epsfysize=5cm
      \epsffile{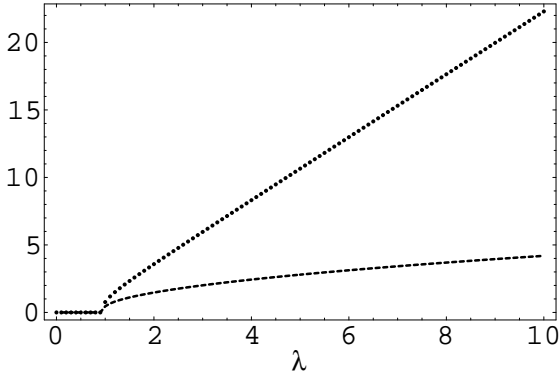}
 \caption{Values for Einstein rings (dotted line) and
for the image separation ($d\beta/d\theta =0$, dashed line),
both for NTS haloes.}
         \label{einst}
  \end{center}
   \end{figure}

\section{Magnification and time delay}

The magnification of lensed images is given by 
\be
\mu = \lt( \frac{\sin{\beta}}{\sin{\theta}} \
\frac{d\beta}{d\theta} \rt)^{-1} \simeq
\lt( \frac{\beta}{\theta} \
\frac{d\beta}{d\theta} \rt)^{-1}\, ,
\ee 
cf.~Narayan \& Bartelmann (1996).
The tangential and radial critical curves (TCC and RCC, respectively)
follow from the singularities of the tangential and the radial magnification
\be
\mu_{\rm t} \equiv \lt(\frac{\sin{\beta}}{\sin{\theta}}\rt)^{-1} \, , \quad
\mu_{\rm r} \equiv \lt(\frac{d\beta}{d\theta}\rt)^{-1} \, ,  
\ee
respectively.

\begin{figure}
\begin{center}
\leavevmode\epsfysize=5cm
      \epsffile{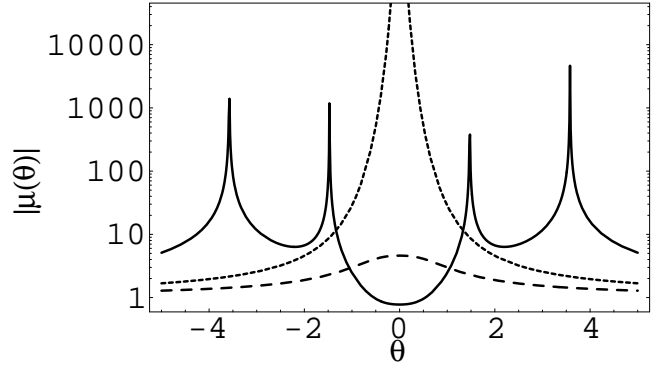}
 \caption{The magnification of a NTS lens.
The solid, dotted, and dashed lines are, respectively,
for $\lambda=2$, $\lambda_{\rm cr}$, $0.3$. The inner peak for $\lambda=2$
notifies the position of a radial critical curve, whereas the
outer peak gives the position of the Einstein ring.
(For the logarithmic representation, the absolute value of
$\mu$ is plotted. The magnifications between the singularities at
both, positive and negative $\theta$, are actually negative.)}
         \label{magn}
  \end{center}
   \end{figure}

Figure \ref{magn} shows the absolute value 
$\vert\mu\vert=\vert(\theta/\beta) (d\theta/d\beta)\vert$
of the magnification of a NTS halo
for several values of $\lambda$.

Let us also determine the Shapiro delay given by
\be
\tau = 2\int \vert \phi\vert  dl =  \frac{\kappa}{4\pi} \int \int
   \frac{\rho ({\vec r}\;^\prime)}{|{\vec r}-
   {\vec r}\;^\prime|}
  d^3 {r}^\prime dl \,.
\ee
It is controlled
dominantly by the Newtonian gravitation potential $\phi$, cf.~Will (2003).

On the other hand, the reduced Shapiro delay $\psi$ is the (super-)potential 
for the reduced deflection angle
\be 
\alpha(\theta) =\nabla_\theta \psi(\theta) =\lambda g(\theta)/\theta \,,
\ee
cf. Eq. (61) of Narayan \& Bartelmann (1996). Thus we find equivalently,
\ba 
\psi(\theta) &=&\lambda \int_0^\theta g(x) d\ln x\nonumber\\
& =&
\frac{\lambda \pi}{8(2-A^2)} \bigl\{ -8- A^2 + 8\sqrt{1+\theta^2} +
\frac{ A^2 }{\sqrt{1+\theta^2}} \nonumber\\
 &-& 
 (8-A^2) \ln[ (1+\sqrt{1+\theta^2})/2 ] \bigr\} \,.
\ea
[In the next order approximation $\widetilde g(x)= g(x) + p(x)$, 
with the radial pressure of the NTS included, we would obtain the additional term
$\lambda \pi A^2(1-1/\sqrt{1+\theta^2})/4(2 -A^2)$.]

\begin{figure}
\begin{center}
\leavevmode\epsfysize=5cm
      \epsffile{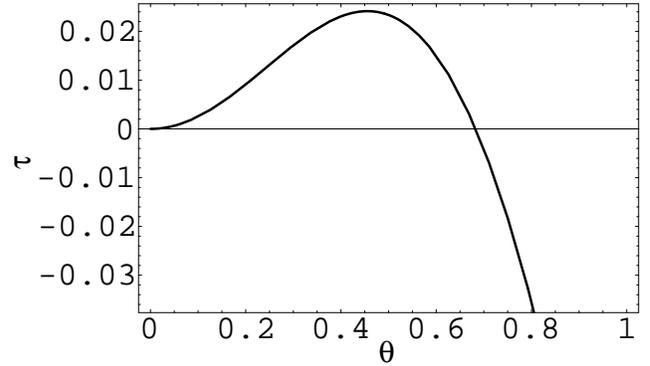}
 \caption{The reduced observed time delay $\widetilde\tau:=(\theta 
-\beta)^2 -2\psi$ as a function of $\theta$ for $A=0.805$ and $\lambda=\lambda_{\rm cr}$.}
         \label{time}
  \end{center}
   \end{figure}

To the observable time delay 
\be 
\Delta t =(1+z_l) \frac{ D_{\rm l}D_{\rm s }}{2D_{\rm ls }}\left[ (\theta 
-\beta)^2 -2\psi\right]
\ee
there contribute  a geometrical part and the Shapiro delay,
where $z_{\rm l}$ is the redshift of the lens.
The result is shown in Fig.~\ref{time}.

\section{Conclusions}

The lens properties of  a solitonic scalar model of DM haloes reveal 
all qualitative features of a non-singular
circularly symmetric gravitational lens. We derived the
lens equation and plotted the corresponding curves, which turn out to be 
quite similar to those of the IS halo.

{}For a comparison with observations, the Castles survey 
(Kochanek, et  al. 2005) of 43 lenses with multiple images
permits us to determine the observed parameter $\lambda$ of the lens equation 
(\ref{dimlens}),  i.e.
\be
\lambda_{\rm obs}=\frac{0.0057}{h} \frac{d_{\rm l} d_{\rm ls}}{d_{\rm s}}
\frac{\rho_0 r_{\rm c}}{{\rm M}_\odot {\rm pc}^{-2}}\,,  \label{lamobs}
\ee 
where $h:=H_0/ 100$ kpc is the normalized Hublle constant and 
$d_{\rm l}:= D_{\rm l}H_0/c$ etc. are dimensionless.
 
The angular diameters have been determined with the code of Kayser et 
al.~(1997) adopting the parameters of the cosmological concordance model.
The scaling relation (\ref{scallaw}) for the sample 
of galaxies studied by Fuchs \& Mielke (2004) implies $\rho_0 r_{\rm c}
\simeq 200$ M$_\odot$pc$^{-2}$, if we use the minimum disk models. We 
find from  (\ref{lamobs}), that 
$\lambda_{\rm obs} 
\leq 0.3$, which implies that no multiple images would occur, 
if the DM haloes were modelled by  the quasi-isothermal sphere (IS), for which 
$\lambda_{\rm cr} =4/\pi$. A similar finding 
was recently deduced from the CSL-1 candidate (Sazhin et al. 2003):  The 
assumption of a DM filament (Fairbairn 2005)
with density $\rho <$ 300 M$_\odot$pc$^{-2}$ is not sufficient in order to 
explain the presumably observed splitting of $8.6\times 10^{-6}$ radiants.

However, for our NTS halo, $\lambda_{\rm cr}$ can approach zero as in the NFW 
case, whereas   haloes modelled by the Burkert profile need an almost three 
times higher $\lambda$ in order to 
produce strong lensing (Park \& Ferguson 2003). This 
differences between the NTS model and the Burkert fit can be
traced back to the fact that the NTS  metric,  as that of the related IS 
profile, corresponds to an
asymptotically constant rotation velocity $v_\infty$.
In a general relativistic setting, 
the space necesarily would 
exhibit a {\em deficit angle} of $4\pi v_\infty^2$ or, equivalently,  
can be joined to a conical metric. Light passing  a NTS halo simulating such a 
`cosmic string' 
will be deflected by this deficit angle 
which, according to (\ref{vinfty}) has the right order of magnitude. 
Consequently, after removing the deficit angle by coordinate transformation, the linear mass density 
peaking at the location of the
central coordinate `singularity' of the conical metric  would
produce a splitting of images also for NTS haloes.

More recently, there are  indications 
of a considerable {\em flattening} of  dark halos from  weak lensing 
observations (Hoekstra et al. 2004) of galaxies with a lower bound 
$e_{\rm halo}= 1-c/a\geq 1/3$ on the  observed halo 
ellipticity.
Our Emden type scalar model has been probed (Mielke \& Peralta, 2004) 
with respect to a   
possible ellipticity of exact NTS solutions, presumably due to rotation with 
angular momentum $l$, and 
round and flattened haloes are found, with an $1/r^{2l+2}$ decay of 
the density at infinity. A  `superposition' of round and elliptic solitons  
to a halo with an effective $l_{\rm eff}=1/2$ could be envisioned which would
have 
the $r^{-3}$ asymptotic
decay of the density, familiar from the 
Burkert and NFW profiles. This idea, however, needs further study. 

With respect to the core/cusp problem 
in low surface brightness galaxies, a new analysis by Gentile et al. (2004) 
has revealed that the Burkert curve -- as any cored profile -- has 
the best fits to the rotation 
curves of a new sample of five spiral galaxies.
In general, the properties of DM haloes inferred from weak lensing
(Hoekstra et al. 2004)
provides a strong support for
the existence of DM, whereas alternative theories of
gravity, such as MOND, can almost be excluded.

\section*{Acknowledgements}
We would like to thank Fjodor Kusmartsev, Humberto Peralta, Remo Ruffini
and Robert Schmidt for helpful discussions and comments.
One of us (F.E.S.) acknowledges
research support provided by a personal fellowship.
Moreover, E.W.M. acknowledges the support of SNI and
thanks  Noelia,  Markus G\'erard Erik, and Miryam Sophie Naomi
for encouragement.

\end{document}